# The pure rotational spectrum of YbOH.

Sanjay Nakhate[a] and Timothy C. Steimle

School of molecular science

Arizona State University

Tempe, Arizona 85287

Nickolas H. Pilgram and Nicholas R. Hutzler

Division of Physics, Mathematics, and Astronomy

California Institute of Technology

Pasadena, California 91125

a) On leave from Atomic and Molecular Physics Division, Bhabha Atomic Research Centre, Mumbai 400 085, India.

**Abstract**

The pure rotational spectrum of YbOH has been recorded and analyzed to produce fine and magnetic hyperfine parameters for the $\tilde{X}\,^2\Sigma^+(0,0,0)$ state. These parameters are compared with those determined from the optical study [Melville and Coxon, *J. Chem. Phys.* **115**, 6974-6978 (2001)] and with the values for YbF [Dickinson *et al.* **115**, 6979-6989 (2001)]. The results support the existence of an unobserved perturbing state near the $\tilde{A}^2\Pi_{1/2}$ state, similar to that previously found in YbF. The precisely determined parameters lay the foundation for laser cooling YbOH, which will aid in the search for new physics beyond the standard model.

**Introduction**

Recently, it has been proposed [1] that the linear triatomic molecule YbOH may be a sensitive venue for investigating T-violating physics beyond the standard model (BSM), such as the electron electric dipole moment (EDM) and nuclear magnetic quadrupole moment (MQM) via the heavy Yb atom. When compared to the diatomic molecules currently used in BSM searches, the additional degrees of freedom offered by YbOH provide several advantages. Specifically, YbOH has the coexistence of an electronic structure amenable to laser cooling, and closely spaced opposite parity states useful for systematic error rejection[1], which are not simultaneously offered by diatomic systems suitable for EDM and MQM searches. A precision measurement with laser cooled and trapped YbOH could extend coherence times (the current limitation of molecular beam BSM searches [2-4]) by orders of magnitude, ultimately probing T-violating BSM physics at the PeV scale[1]. The analogous study of laser cooled YbF has been actively pursued for some time [2, 3, 5, 6]. YbOH offers several advantages compared to YbF. First, the metastable (0,1,0) bending mode of YbOH contains closely spaced states of opposite parity, which enables full polarization of the molecule in the lab frame as well as internal co-magnetometry[1]. Internal co-magnetometers provide natural systematic error rejection, something relied upon by the most sensitive searches for the electron EDM[7, 8]. Additionally, YbOH has reduced proton magnetic hyperfine splitting of the optical transitions, as compared to that of $^{19}$F in YbF, which simplifies the experimental approach for laser cooling.

Effective laser cooling of a molecular species require many photon scattering events, and therefore all of the transitions from the hyperfine manifold of the ground state must be addressed. This is accomplished by applying radio frequency sidebands to the main and vibrational repump lasers, and has been utilized to cool and trap diatomic molecules[9-11] as well as cool polyatomic molecules[12]. The spin-rotation and hyperfine splitting of the ground state must be known accurately to address these multiple internal levels. An objective of the current study is to precisely determine the spin rotation and proton hyperfine splitting in the ground electronic state.

Previous spectroscopic studies of YbOH were limited to the high temperature, Doppler limited, optical study of the 000–000 and 000–100 bands of the $\tilde{A}\,^2\Pi_{1/2} - \tilde{X}\,^2\Sigma^+$ electronic transition[13]. Here, the low temperature, low-lying, pure rotational spectrum of the $\tilde{X}\,^2\Sigma^+(0,0,0)$ ground state of YbOH is presented. The spectrum was recorded using a pump/probe microwave optical double resonance technique on a cold supersonic molecular beam. The resulting rotational spectrum was analyzed to precisely determine the rotational, spin-rotational, and hyperfine parameters of the $\tilde{X}\,^2\Sigma^+(0,0,0)$ state. The determined spectroscopic parameters are compared with those determined in previous studies and the corresponding parameters for YbF[14] to garner insight into excited electronic states.

**Experimental**

The experimental set-up for separated field, pump/probe microwave optical double resonance (PPMODR) measurements has been described previously [15, 16]. Briefly, a pulsed supersonic molecular beam sample of YbOH was produced by skimming the output of an ablated Yb rod in the presence of a methanol/argon expansion. The sample was sequentially exposed to an intense (~100 mW) pump beam that removes population from a state of interest, tunable microwave radiation that repopulates the state, and a weak (~5 mW) probe laser beam. Microwave transitions were detected as an increase in the laser induced fluorescence monitored at the probe laser beam region. The fluorescent signal resulting from the cw-dye laser excitation of the 000–000 bands of the $\tilde{A}\,^2\Pi_{1/2} - \tilde{X}\,^2\Sigma^+$ electronic transition was monitored on resonance and processed using gated photon counting techniques. Active harmonic multipliers, in conjunction with a 0-20 GHz synthesizer, were used for generating the tunable microwave radiation. A rubidium standard was used for the time-base of the synthesizer.

**Observations**

Rotational transitions involving the $f_1$-components (i.e. $J=N+1/2$) were detected by pumping and probing the $Q_1$-branch features, whereas those involving the $f_2$-components (i.e. $J=N-1/2$) were detected by pumping and probing $Q_{12}$ or $R_{12}$ branch features. Complete analysis of the 000–000 $\tilde{A}\,^2\Pi_{1/2} - \tilde{X}\,^2\Sigma^+$ molecular beam optical spectrum will be presented elsewhere. As an example, the overlapped spectrum for the $N''=5, J''=11/2, F''=5$ and $6 \to N''=6, J''=13/2, F''=6$ and 7 transitions is given in Figure 1. Also presented in Figure 1 are the associated energy levels and assignments. The signal was recorded using approximately 1 µW of microwave radiation and becomes broader with increasing power. The radiation was stepped in increments of 6 kHz and the LIF photon counts of approximately 600 laser ablation shots were summed at each frequency step. Twelve measured spectral features were assigned to 18 transition frequencies of the $\tilde{X}\,^2\Sigma^+(0,0,0)$ state. The observed and calculated transition frequencies, the differences between them, and the associated quantum number assignments are presented in Table 1.

Figure 1

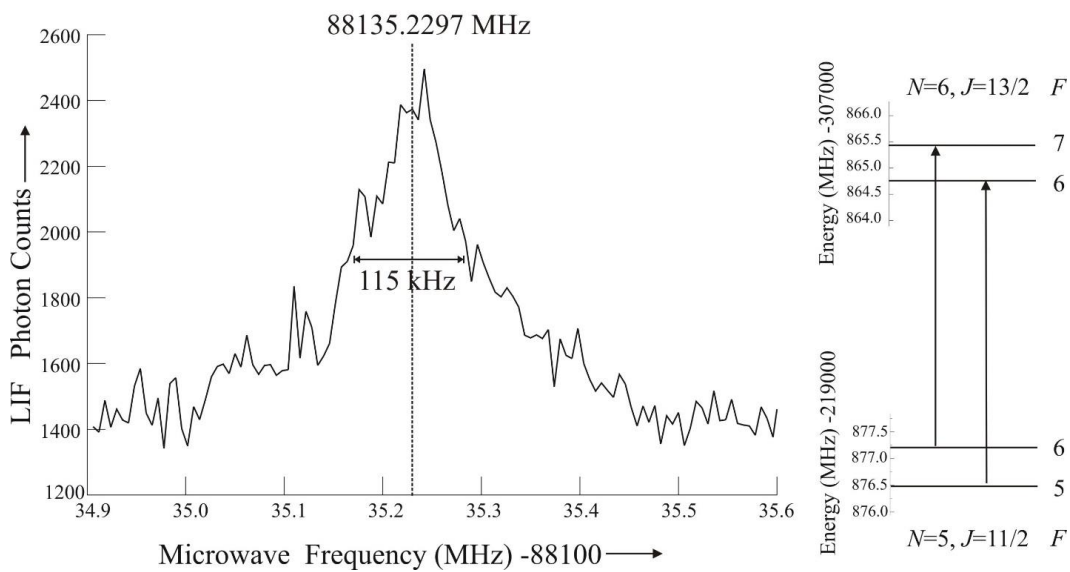

**Figure 1.** The overlapped $N''=5\ J''=11/2,\ F''=5$ and $6 \rightarrow N''=6\ J''=13/2,\ F''=6$ and 7 transitions and associated energy level assignments.

Table 1. Rotational transitions in the $\tilde{X}\,^2\Sigma^+(0,0,0)$ of YbOH

| N", J", F" | N', J', F' | Obs. (MHz) | Calc. (MHz) | Obs.-Calc.(MHz) |
|---|---|---|---|---|
| 3, 7/2, 4 | 4, 9/2, 5 | 58745.1654 | 58745.1640 | 0.0014 |
| 3, 7/2, 3 | 4, 9/2, 4 | 58745.2594 | 58745.2642 | -0.0048 |
| 3, 5/2, 3 | 4, 7/2, 4 | 58826.0166 | 58826.0104 | 0.0062 |
| 3, 5/2, 2 | 4, 7/2, 3 | 58826.1460 | 58826.1334 | 0.0126 |
| 3, 5/2, 3 | 4, 7/2, 3 | 58827.5935 | 58827.6040 | -0.0105 |
| 4, 9/2, 5 | 5, 11/2, 5 | 73438.2470 | 73438.2575 | -0.0105 |
| 4, 9/2, 5 | 5, 11/2, 6 | 73440.5890 | 73440.5464 | 0.0426 |
| 4, 9/2, 4 | 5, 11/2, 5 | 73440.5890 | 73440.6109 | -0.0219 |
| 4, 7/2, 3 | 4, 9/2, 4 | 73521.3229 | 73521.3677 | -0.0448 |
| 4, 7/2, 4 | 4, 9/2, 5 | 73521.3229 | 73521.2922 | 0.0307 |
| 5, 11/2, 6 | 6, 13/2, 7 | 88135.2297 | 88135.2115 | 0.0182 |
| 5, 11/2, 5 | 6, 13/2, 6 | 88135.2297 | 88135.2565 | -0.0268 |
| 5, 9/2, 5 | 6, 11/2, 6 | 88215.8438 | 88215.8150 | 0.0288 |
| 5, 9/2, 4 | 6, 11/2, 5 | 88215.8438 | 88215.8662 | -0.0224 |
| 6, 13/2, 7 | 7, 15/2, 8 | 102829.0317 | 102829.0140 | 0.0177 |
| 6, 13/2, 6 | 7, 15/2, 7 | 102829.0317 | 102829.0472 | -0.0155 |
| 6, 11/2, 6 | 7, 13/2, 6 | 102909.4589 | 102909.4405 | 0.0184 |
| 6, 11/2, 5 | 7, 13/2, 5 | 102909.4589 | 102909.4776 | -0.0187 |

**Analysis**

The effective Hamiltonian used to model the energy is [17]:

$$\mathbf{H}^{eff}(^2\Sigma^+) = B\mathbf{N}^2 - D(\mathbf{N}^2)^2 + \gamma \mathbf{N}\cdot\mathbf{S} + \gamma_D\left[\mathbf{N}\cdot\mathbf{S}, \mathbf{N}^2\right]_+ + b_F \mathbf{I}\cdot\mathbf{S} + \frac{c}{3}(3I_z S_z - \mathbf{I}\cdot\mathbf{S}), \quad (1)$$

where **N** is the angular momentum operator excluding electronic and nuclear spin, **S** and **I**, respectively and []$_+$ is an anti-commutator. Eigenvalues and eigenvectors were generated by constructing and diagonalizing a 4 x 4, Hund's case a$_{\beta J}$ matrix representation. The expressions for the matrix elements were taken from the literature [17]. The eigenvalues and observed transition wave numbers were input to a non-linear least-squares fitting algorithm to determine the spectroscopic parameters of the $\tilde{X}\,^2\Sigma^+(0,0,0)$ state. The results of the fit are presented in Table 2 along with the determined parameters from the previous optical analysis [13]. The standard deviation of the fit was 23 kHz, which is commensurate with the estimated measurement uncertainty. Also presented are the spectroscopic parameters of the ground state of $^{174}$YbF derived from the analysis of the Fourier transform microwave absorption spectrum[14].

Table 2. Spectroscopic constants for the $\tilde{X}\,^2\Sigma^+(0,0,0)$ state of $^{174}$YbOH (MHz)

| Parameter | Values[a] | | YbF(v=0)[c] |
|---|---|---|---|
| | Present | Previous[b] | |
| $B$ | 7348.40053(29) | 7357.92(39) | 7233.827(17) |
| $D$ | 0.006084(39) | 0.006535(84) | 0.007159(fixed) |
| $\gamma$ | -81.150(57) | 28.90(42) | -13.41679(13) |
| $\gamma_D$ | 0.00476(56) | | 0.0039840(15) |
| $b_F$ | 4.80(18) | | 170.26374(20) |
| $c$ | 2.46(48) | | 85.4028(19) |

a) The numbers in parenthesis represents a $2\sigma$ error estimate.
b) Reference 13
c) Reference 17

**Discussion**

The determined rotational $B$ (7348.40053(29) MHz) and centrifugal distortion $D$ (0.006084(39) MHz) parameters are similar to those derived from the analysis of the optical spectrum[13] but much more precisely determined. Making the reasonable assumption that the O-H bond distance is that for ground state BaOH(0.9270Å)[18], the $B$ value gives a Yb-O distance of 2.0397 Å, which is nearly identical to the 2.0165 Å bond distance of $^{174}$YbF[14]. The newly determined spin-rotation parameter, $\gamma$ (-81.150(57) MHz), is approximately three times larger in magnitude and of opposite sign from that derived from the previous optical analysis [13]. It is noteworthy that the dataset for the previous high temperature optical study was dominated by high rotational levels and not those probed in the present cold molecular beam study.

Insight into the nature of the excited electronic states of YbOH, which is relevant for the proposed optical cooling [1], can be derived from the interpretation of $\gamma$ and a comparison with $^{174}$YbF. This parameter is dominated by second order effects:

$$\gamma^{(2)} = 2\sum_{^2\Pi} \frac{\left\langle ^2\Sigma_{-1/2} \left| BL^- \right| ^2\Pi_{1/2} \right\rangle \left\langle ^2\Pi_{1/2} \left| \sum_i a_i l_i^+ s_i^- \right| ^2\Sigma_{1/2} \right\rangle}{T(^2\Pi) - T(^2\Sigma)}. \tag{2}$$

The determined γ values for YbF($X\,^2\Sigma^+$) are -13.41679 [14], -33.81036[14], -53.8[19] and -74.5 MHz [19] for v=0, 1, 2 and 3 respectively. This observed strong vibrational dependence for YbF was attributed to competing second-order contributions to γ from the $A\,^2\Pi_{1/2}$ and another nearby excited Ω=1/2 state [19]. The analysis of the (0,0) $X\,^2\Sigma^+$–$A^2\Pi_{1/2}$ band [20] of YbF performed long ago also concluded that more than one state enters into the summation of Eq. 2 for YbF. The $A\,^2\Pi_{1/2}(v=0)$ ($T$=18107 cm$^{-1}$), [561]($T$= 18581 cm$^{-1}$) and [557] ($T$= 18706 cm$^{-1}$) excited states of YbF are most likely the dominant contributors. The primary configuration for the $A\,^2\Pi_{1/2}(v=0)$ has a single unpaired electron in a Yb$^+$-centered π-type orbital and makes a positive contribution to $\gamma^{(2)}$. The [561] and [557] excited states of YbF have a large contribution from a configuration which has a hole in the 4f core. This configuration makes a negative contribution to $\gamma^{(2)}$. These two configurations should also dominate the description of the low-lying excited states of isovalent YbOH. Like YbF($X\,^2\Sigma^+$), γ for YbOH is negative but approximately six times larger, suggesting that the 4f$^{13}$ configuration is more important for the description of the low-lying excited states. Also noteworthy is that $\gamma_D$(0.00476(56) MHz) for YbOH is extremely close to that of YbF, $\gamma_D$(=0.0039840(15) MHz). This relatively large centrifugal distortion causes spin-rotation splitting to change sign at higher rotational levels and could account for the difference between the newly determined value of γ and that from the previous work [13].

The magnetic hyperfine parameters, $b_F$(H)(= 4.80(18) MHz) and $c$(H) (= 2.46(48) MHz) are relatively small and have similar values as those determined for the $\tilde{X}\,^2\Sigma^+(0,0,0)$ state of SrOH[21], $b_F$(H) (=1.713(2) MHz) and $c$(H)(=1.673(5) MHz). As expected the determined hyperfine parameters are much smaller than those of YbF, $b_F$(F)(= 170.26374(20) MHz) and $c$(F) (= 85.4028(19) MHz) because H is farther removed from the Yb centered unpaired electron. This small splitting indicates that the hyperfine levels of YbOH will not need to be individually addressed when laser cooling YbOH.

**Conclusion**

The spectroscopic parameters for the $\tilde{X}\,^2\Sigma^+(0,0,0)$ state of YbOH have been precisely determined. The data will be useful in the design of proposed cooling and trapping experiments. The large negative $\gamma$ indicates that the excited electronic states arising from the $4f^{13}$ configuration will play an important role in the proposed laser-cooling scheme [1]. The determined bond lengths and magnetic hyperfine parameters will be useful for assessing computational methodologies for precisely predicting effective internal electric fields and other properties relevant to fundamental physics. Generation of an intense supersonic molecular beam of YbOH via a laser ablation scheme has been demonstrated.

**Acknowledgment**

The research at Arizona State University and Caltech was supported by a grant from the Heising-Simons Foundation (Grant 2018-0681).